\documentclass[12pt]{article}
\usepackage{amsmath, amssymb}
\usepackage{url}
\usepackage{graphicx}

\begin{document}
\title{Acceleration-Induced Nonlocal Electrodynamics in
Minkowski Spacetime}

\author{Uwe Muench\\Friedrich W. Hehl\thanks{Permanent address:
Inst.\ Theor.\ Physics, Univ.\ of Cologne, 50923 K\"oln,
Germany}\\Bahram Mashhoon} 

\date{Department of Physics \& Astronomy, University of Missouri,\\
Columbia, MO 65211, USA}
\maketitle

\begin{abstract} 
We discuss two nonlocal models of electrodynamics in which the
nonlocality is induced by the acceleration of the observer. Such an
observer actually measures an electromagnetic field that exhibits
persistent memory effects. We compare Mashhoon's model with a new
ansatz developed here in the framework of charge \& flux
electrodynamics with a constitutive law involving the Levi-Civita 
connection as seen from the observer's local frame and conclude that
they are in partial agreement only for the case of constant
acceleration.  
\hfill {\em Files kernel14.tex +
kernel14a.ps + kernel14a.pst + kernel14b.ps + kernel14b.pst +
kernel14c.ps, 2000-03-23}
\end{abstract}

\section{Introduction}

As pointed out by Einstein \cite{Einstein}, in special relativity
theory it is assumed that the rate of a fundamental (``ideal'') clock
depends on its instantaneous speed and is not affected by its
instantaneous acceleration. This is usually called the ``clock
hypothesis''; see \cite{Stedman,Stedman2,Bahrama} for more recent
discussions of this assumption. The decay of elementary particles
obeys this hypothesis very well as shown by Eisele \cite{Eisele} for
the weak decay of the muon.

If we study electrodynamics, for instance, in an accelerated reference
frame (see \cite{Heintz}), then we have to presuppose corresponding
hypotheses for the measurement of the electric and magnetic fields,
the electric charge, etc. In this way, we arrive at the {\em hypothesis of
locality} that has been extensively investigated
\cite{Bahram3,Bahram1,Bahram93,BahramBras,Bahram2}.  
Replacing the curved worldline of the accelerated observer by its
instantaneous tangent vector is reasonable if the intrinsic spacetime
scales of the phenomena under consideration are negligibly small
compared to the characteristic acceleration scales that determine the
curvature of the worldline; otherwise, the past worldline of the
observer must be taken into account. This would then result in a
nonlocal electrodynamics for accelerated systems.

Nonlocal constitutive relations have been studied in the
\emph{phenomenological} electrodynamics of continuous media for a
long time \cite{Landau,Eringen}. In basic field theories, form-factor
nonlocality has been the subject of extensive investigations. 
The main problem with such field theoretical approaches has been that they
defy quantization. A review of nonlocal quantum field theories and
their insurmountable difficulties has been given by Marnelius
\cite{Marnelius}. The present work is concerned with a benign form of
nonlocality that is induced by the acceleration of the observer.

The hypothesis of locality refers directly to {\em
acceleration}; therefore, one can develop an alternative approach in
which the acceleration enters as the decisive quantity. This type of 
nonlocality, if it refers to time, would involve persistent memory
effects. Materials with memory have been extensively studied. However,
we are interested in the ``material'' vacuum -- and in 
this context our paper is devoted to a comparison of two models
involving acceleration-induced nonlocality. 

\section{Mashhoon's model}

The observational basis of the special theory of relativity generally 
involves measuring devices that are accelerated; for instance, static
laboratory devices on the Earth participate in its proper
rotation. The standard extension of Lorentz invariance to accelerated
observers in Minkowski spacetime is based on the hypothesis of
locality, namely, the assumption that an accelerated observer is
locally equivalent to a momentarily comoving inertial observer. The
worldline of an 
accelerated observer in Minkowski spacetime is curved and this
curvature depends on the observer's translational and rotational
acceleration scales. The hypothesis of locality is thus reasonable if
the curvature of the worldline could be ignored, i.e.\ if the
phenomena under consideration have intrinsic scales that are
negligible as compared to the acceleration scales of the observer. The
accelerated observer passes through a continuous infinity of
hypothetical comoving inertial observers along its worldline;
therefore, to go beyond the hypothesis of locality, it appears natural
to relate the measurements of an accelerated observer to the class of
instantaneous comoving inertial observers.

Consider, for instance, an electromagnetic radiation field $F_{ij}$ in
an inertial frame and an accelerated observer carrying an {\em
orthonormal tetrad frame} $e^i{} _{\!\alpha}(\tau)$ along its
worldline. Here $\tau$ is its proper time, 
the Latin indices $i$, $j$, $k$, \dots{}, which run from 0 to 3, refer
to spacetime coordinates (holonomic indices), while the Greek indices
$\alpha$, $\beta$, $\gamma$, \dots{}, which run from $\hat{0}$ to
$\hat{3}$, refer to (anholonomic) frame indices, and we
choose the signature $(+,-,-,-)$. The hypothesis of locality implies
that the field as measured by the observer is the projection of
$F_{ij}$ upon the frame of the instantaneously comoving inertial
observer, i.e.
\begin{equation}\label{field'}
F_{\alpha\beta}(\tau)= F_{ij}\,e^i{} _{\!\alpha}\,
e^j{}_{\!\beta}\,.
\end{equation}
On the other hand, measuring the properties of the radiation field
would necessitate finite intervals of time and space that would then
involve the curvature of the worldline. The most general {\em
linear} relationship between the measurements of the accelerated
observer and the class of comoving inertial observers consistent with
causality is
\begin{equation}\label{non-local1}
{\cal F}_{\alpha\beta}(\tau) =
F_{\alpha\beta}(\tau)+\int\limits_{\tau_0}^\tau
K_{\alpha\beta}{}^{\gamma\delta}(\tau,\tau')\,
F_{\gamma\delta}(\tau')\,d\tau'\,,
\end{equation}
where ${\cal F}_{\alpha\beta}$ is the \emph{field actually measured},
$\tau_0$ is the instant at which the acceleration begins and the
kernel $K$ is expected to depend on the acceleration of the
observer. A nonlocal theory of accelerated observers has been
developed \cite{Bahram93,BahramBras} based on the assumptions that (i)
$K$ is a convolution-type kernel, i.e.\ it depends only on
$\tau-\tau'$, and (ii) the radiation field never stands completely
still with respect to an accelerated observer. The latter is a
generalization of a consequence of Lorentz invariance for inertial
observers to all observers.

In the space of continuous functions, the Volterra integral equation
(\ref{non-local1}) provides a unique relationship between ${\cal
F}_{\alpha\beta}$ and $F_{\alpha\beta}$. It is possible to
express (\ref{non-local1}) as
\begin{equation}\label{non-local2}
F_{\alpha\beta}(\tau)={\cal F}_{\alpha\beta}(\tau)
+\int\limits_{\tau_0}^\tau
R_{\alpha\beta}{}^{\gamma\delta}(\tau,\tau')\, {\cal
F}_{\gamma\delta}(\tau')\,d\tau'\,,
\end{equation}
where $R$ is the resolvent kernel and if $K$ is a convolution-type
kernel as we have assumed in (i), then so is $R$, i.e.\
$R=R(\tau-\tau')$. Assumption (ii) then implies that
\begin{equation}\label{resolvent}
R(\tau)=\frac{d\Lambda(\tau+\tau_0)}{d\tau}\,\Lambda^{-1}(\tau_0)\,,
\end{equation}
where $R$ and $\Lambda$ are $6\times 6$ matrices and $\Lambda$ is
defined by (\ref{field'}) expressed as $\hat{F}=\Lambda\, F$ in the
six-vector notation. Here $\hat{F}$ denotes the field as referred to the
anholonomic frame. This nonlocal theory, which is consistent with all
observational data available at present, has been described in detail
elsewhere \cite{BahramBras}.

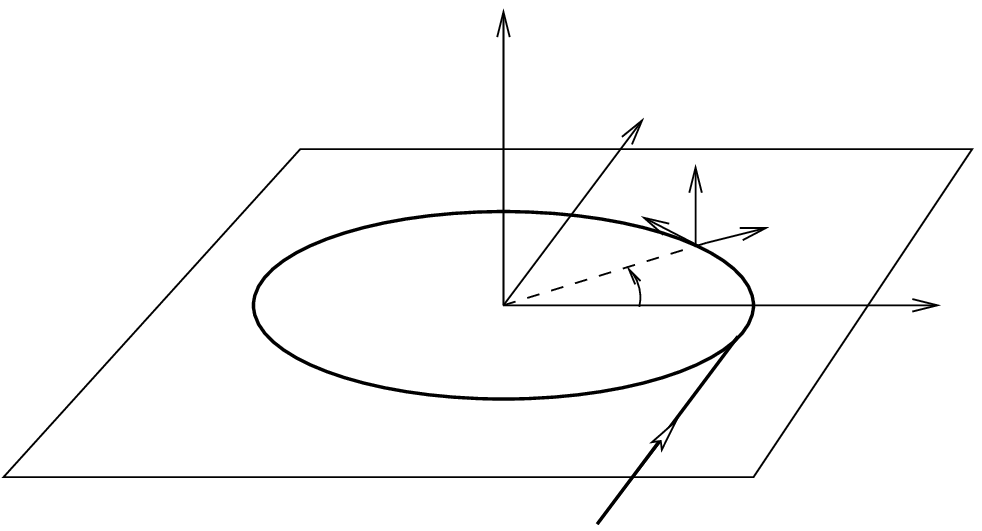 
\bigskip

\noindent Fig.1. The path of an observer in space moving with
constant angular velocity  around the $z$-axis for $\tau >
\tau_0$.
\bigskip 
 
It proves interesting to provide a concrete example of the nonlocal
relationship (\ref{non-local1}). Imagine an observer that moves
uniformly in the inertial frame along the $y$-axis with speed $c\beta$
for $\tau <\tau_0$ and for $\tau\geq \tau_0$ rotates with uniform
angular speed $\Omega$ about the $z$-axis on a circle of radius $r$,
$\beta=r\,\Omega/c$, in the $(x,y)$-plane, see Fig.1. In this case,
\begin{eqnarray}\label{rotframe}
e^i{}_{\!\hat{0}}
&=&\gamma\,\,(1,\>-\beta\sin\varphi,\>\beta\cos\varphi,\>0)\,,\nonumber\\
e^i{}_{\!\hat{1}}&=&\quad(0,\>\>\>\>\>\>\,\cos\varphi,\>\>\>\>\>
\sin\varphi,\>0)\,,\\
e^i{}_{\!\hat{2}}&=&\gamma\,\,(\beta,\>\>\>-\sin\varphi,\>\>\>\>
\cos\varphi,\>0)\,, \nonumber\\
e^i{}_{\!\hat{3}}&=&\quad(0,\quad\qquad 0,\!\quad\qquad 0,\>
1)\,,\nonumber
\end{eqnarray}
in $(ct,x,y,z)$ coordinates with
$\varphi=\Omega(t-t_0)=\gamma\,\Omega(\tau-\tau_0)$. Here $\varphi$ is
the azimuthal angle in the $(x,y)$-plane and $\gamma$ is the Lorentz
factor. Using six-vector notation,
\begin{equation}\label{6vector}
(F_{\alpha\beta}
)\rightarrow
\begin{bmatrix}
\boldsymbol{\hat{E}} \\
\boldsymbol{\hat{B}}
\end{bmatrix}\,,\qquad\quad
({\cal F}_{\alpha\beta})\rightarrow
\begin{bmatrix}
\boldsymbol{\cal {E}} \\
\boldsymbol{\cal {B}}
\end{bmatrix}\,,
\end{equation}
one can show that with respect to the tetrad frame
(\ref{rotframe})
\begin{eqnarray}\label{explicit1}
\boldsymbol{\cal
E}&=&\boldsymbol{\hat{E}}+\int\limits_{\tau_0}^\tau\left[
\boldsymbol{\omega}\times\boldsymbol{\hat{E}}(\tau')
-\frac{\boldsymbol{a}}{c}\times\boldsymbol{\hat{B}}(\tau')\right]\,d\tau'\,,\\
\label{explicit2} \boldsymbol{\cal
B}&=&\boldsymbol{\hat{B}}+\int\limits_{\tau_0}^\tau\left[
\frac{\boldsymbol{a}}{c}\times\boldsymbol{\hat{E}}(\tau')
+\boldsymbol{\omega}\times\boldsymbol{\hat{B}}(\tau')\right]\,d\tau'\,,
\end{eqnarray}
where $\boldsymbol{a}$ is the constant centripetal acceleration of the
observer and $\boldsymbol{\omega}$ is its constant angular
velocity. These quantities can be expressed with respect to the triad
$e^i{}_{\!A}$ as $\boldsymbol{a}=(-c\beta\gamma^2\,\Omega,\,0,\,0)$
and $\boldsymbol{\omega}=(0,\,0,\,\gamma^2\,\Omega)$. For an arbitrary
accelerated observer, we expect that the relations analogous to
(\ref{explicit1}) and (\ref{explicit2}) would be much more
complicated.

Imagine now a general congruence of accelerated observers such that
relations similar to (\ref{non-local1}) and (\ref{non-local2}) hold
for each member of the congruence. The requirement that the
electromagnetic field $F_{ij}$ (or $F_{\alpha\beta}$) satisfy
Maxwell's equations would then imply, via (\ref{non-local2}), that the
field ${\cal F}_{\alpha\beta}$ would satisfy certain complicated
integro-differential equations, which could then be regarded as the
nonlocal Maxwell equations for ${\cal F}_{\alpha\beta}$. Instead of
this system, we give here a different, but analogous,
acceleration-induced nonlocal electrodynamics and study some of its
main properties.

\section{Charge \& flux electrodynamics with a new nonlocal ansatz}

The electrodynamics of charged particles and flux lines, see
\cite{Honnef2,Obukhov} and the references cited therein,
involves the electromagnetic field strength $F_{\alpha\beta}$---that
is defined via the Lorentz force law and is directly related to the
conservation law of magnetic flux---as well as the electromagnetic
excitation $\mathcal{H}^{\alpha\beta}$ that is directly related to the
electric charge conservation. The corresponding Maxwell
equations are metric-free and in Ricci calculus in arbitary 
frames read (cf.\ \cite{schouten,Post})
\begin{eqnarray}\label{Maxwell}
 \partial_{[\alpha}\,F_{\beta\gamma]}
-C_{[\alpha\beta}{}^\delta\,F_{\gamma ]\delta} &=& 0\,,\\
\partial_\beta\,{\cal
H}^{\alpha\beta}-\frac{1}{2}\,C_{\beta\gamma}{}^\alpha\,{\cal
H}^{\gamma\beta}-\frac{1}{2}\,C_{\beta\gamma}{}^\beta\,{\cal
H}^{\alpha\gamma}& =& {\cal J}^\alpha\,.
\end{eqnarray} Here $\mathcal{J}^\alpha$ is the electric current and
the $C$'s are the components of the object of anholonomicity:
\begin{equation}\label{anholonomicity}
C_{\alpha\beta}{}^\gamma:=2\,e^i{}_{\!\alpha}e^j{}_{\!\beta}
\,\partial_{[i}\,e_{j]}{}^{\!\gamma}=- C_{\beta\alpha}{}^\gamma\,.
\end{equation}
Ordinarily for vacuum, we would have the constitutive equation 
\begin{equation}\label{vacuum}
{\cal H}^{\alpha\beta}=\sqrt{-g}\,g^{\alpha\mu}
\,g^{\beta\nu}\,F_{\mu\nu}\,.
\end{equation}
However, this reformulation of electrodynamics allows for much more
general constitutive relations between $\mathcal{H}^{\alpha\beta}$ and
$F_{\alpha\beta}$. In particular, it is possible to develop a nonlocal
\emph{ansatz} based on a generalization of \eqref{vacuum} along the
lines suggested by Obukhov and Hehl  \cite{Honnef2}
\begin{equation}\label{kernel-basic}
\mathcal{H}^{\alpha\beta}(\tau,\xi)=\sqrt{-g}\,g^{\alpha\mu}
\,g^{\beta\nu} \int \mathcal{K}_{\mu\nu}{}^{\rho\sigma} (\tau, \tau', \xi)
F_{\rho\sigma}(\tau',\xi) \, d\tau'\;,
\end{equation}
where the kernel $\mathcal{K}$ corresponds to the response of the
medium and $\xi^A$, $A=1,2,3$, are the Lagrange coordinates of the
medium.

As an alternative to Mashhoon's model but along the same line of
thought, see equation \eqref{non-local1}, one can develop an 
acceleration-induced nonlocal constitutive relation in vacuum via
equation \eqref{kernel-basic} by using the ansatz,
\begin{eqnarray}\label{kernel}
{\cal H}^{\alpha\beta}(\tau)&=&\sqrt{-g}\,g^{\alpha\mu}
\,g^{\beta\nu}\Big[F_{\mu\nu}(\tau) \Big.\nonumber\\ &&
- c \int\limits_{\tau_0}^{\tau}[\Gamma_{0\mu}{}^{\rho}
(\tau-\tau')F_{\rho\nu}(\tau')+ \Big.\Gamma_{0\nu}{}^\rho
(\tau-\tau')F_{\mu\rho}(\tau')]\,d\tau '\Big]\,,
\end{eqnarray} 
where the integral is over the worldline of an accelerated observer in
Min\-kowski spacetime as before. Here the response of the ``medium'' is
simply given by the Levi-Civita connection of the accelerated observer
in vacuum and the local constitutive relation \eqref{vacuum} is recovered for
\emph{inertial} observers. 

We recall that in an {\em orthonormal} frame the connection is
equivalent to the anholonomicity, see \cite{schouten}:
\begin{equation}\label{conn}
\Gamma_{\alpha\beta\gamma}:=g_{\gamma\delta}\,\Gamma_{\alpha\beta}{}^\delta
  =\frac{1}{2}\,(-C_{\alpha\beta\gamma}+C_{\beta\gamma\alpha}
  -C_{\gamma\alpha\beta})=-\Gamma_{\alpha\gamma\beta}\,.
\end{equation}
If we invert (\ref{conn}), we find that
$C_{\alpha\beta\gamma}=-2\Gamma_{[\alpha\beta]\gamma}$.

In the following, we explore the consequences of the new ansatz
\eqref{kernel} for a general accelerated observer in Minkowski
spacetime. 

\section{The new ansatz and the accelerating and rotating 
observer}

It has been shown in \cite{WeiTou,Honnef1}, and the references cited
therein, that the {\em orthonormal} frame $e_\alpha$ of an arbitrary 
observer with local 3-acceleration $\boldsymbol{a}$ and local
3-angular velocity $\boldsymbol{\omega}$ reads
\begin{eqnarray}\label{frame}
e_{\hat 0} &=& \frac{1}{1+\frac{\boldsymbol{a}}
  {c^2}\cdot\overline{\boldsymbol{x}}}
  \left[\partial_{\overline{0}}-\left( \frac{\boldsymbol{\omega}}{c}
  \times{\overline{\boldsymbol{x}}} \right)
  ^{\overline{B}}\partial_{\overline{B}}\right]\,,\nonumber\\
  e_{A}&=&\partial_{\overline{A}}\,,
\end{eqnarray}
where the barred coordinates are the standard normal coordinates
adapted to the worldline of the accelerated observer.  The coframe
$\vartheta^\alpha$ can be computed by inversion. We find
\begin{eqnarray}\label{coframe}
\vartheta^{\hat 0} &=&  \nonumber 
\left(1+\frac{\boldsymbol{a}}{c^2}\cdot\overline{\boldsymbol{x}}\right)\,
    dx^{\overline{0}} =Ndx^{\overline{0}}\,,\\ \vartheta^{A} &=
    &dx^{\overline{A}} +\left(\frac{\boldsymbol{\omega}}{c}
    \times\overline{\boldsymbol{x}} \right)^{\overline{A}}\,
    dx^{\overline{0}} =dx^{\overline{A}}+N^{\overline{A}}
    dx^{\overline{0}}\,.
\end{eqnarray}
In the $(1+3)$-decomposition of spacetime, $N$ and $N^{\overline{A}}$
are known as {\em lapse function} and {\em shift vector},
respectively. The frame and the coframe are orthonormal. The
metric reads as follows:
\begin{eqnarray}\label{metric}
ds^2=\eta_{\alpha\beta}\,\vartheta^\alpha\otimes\vartheta^\beta
 &=& \nonumber \left[\left(1+\frac{\boldsymbol{a}}{c^2}\cdot
 \overline{\boldsymbol{x}}\right)^2
-\left(\frac{\boldsymbol{\omega}}{c}\times\overline{\boldsymbol{x}}\right)^2
 \right]\,\left(dx^{\overline{0}}\right)^2\\ & &
 -2\,\left(\frac{\boldsymbol{\omega}}{c}
 \times{\overline{\boldsymbol{x}}}\right)_{\overline{A}}\,
 dx^{\overline{0}}\,
 dx^{\overline{A}}-\delta_{{\overline{A}}\,{\overline{B}}}dx^{\overline{A}}
 dx^{\overline{B}}\,,
\end{eqnarray}
where $(\boldsymbol{\omega}
\times{\overline{\boldsymbol{x}}})_{\overline{A}}
=\epsilon_{{\overline{A}}\,{\overline{B}}\,
{\overline{C}}}\,\omega^{\overline{B}}\,x^{\overline{C}}$, 
$\boldsymbol{a}=a^{\overline{A}}\,e_{\overline{A}}$, and
$a^{\overline{A}}=e_i{}^{\overline{A}}\,a^i$.

Starting with the coframe, we can read off the connection coefficients
(for vanishing torsion) by using Cartan's first structure equation
$d\vartheta^\alpha=-\Gamma_\beta{}^\alpha\wedge\vartheta^\beta$
with $\Gamma_\beta{}^\alpha= \Gamma_{{\overline{i}} \beta}{}^\alpha\,
dx^{\overline{i}}$.  By construction, the connection projected in
spacelike directions vanishes, since we have spatial Cartesian 
laboratory coordinates. Thus we are left with the following
nonvanishing connection coefficients:
\begin{eqnarray}\label{conn1}
\Gamma_{{\overline{0}}{\hat 0}A}&=&\nonumber -\Gamma_{{\overline{0}} A{\hat 0}}
  ={\frac{a_A}{ c^2}}\,,\\ \Gamma_{{\overline{0}}
  AB}&=&-\Gamma_{{\overline{0}} BA}
  =\epsilon_{ABC}\,{\frac{\omega^C}{c}}\,.
\end{eqnarray}
The first index in $\Gamma$ is holonomic, whereas the second and third
  indices are 
  anholonomic. If we transform the first index, by means of the frame
  coefficients $e^{\overline{i}}{}_\alpha$, into an anholonomic one,
  then we find the totally anholonomic connection coefficients as
  follows:
\begin{eqnarray}\label{conn2}
\Gamma_{{\hat 0}{\hat 0}A} &=& -\Gamma_{{\hat 0}A{\hat 0}} =
  \frac{{a_A}/{c^2}}{1+\boldsymbol{
  a}\cdot{\overline{\boldsymbol{x}}}/c^2}\,,\nonumber\\ \Gamma_{{\hat
  0}AB}&=&-\Gamma_{{\hat 0}BA}
  =\frac{\epsilon_{ABC}\,\omega^C/c}{ 1+\boldsymbol{a}
  \cdot{\overline{\boldsymbol{x}}}/c^2}\,.
\end{eqnarray}
In general, of course, the translational acceleration $\boldsymbol{a}$
and the angular velocity $\boldsymbol{\omega}$ are {\em functions of
time}.

Let us return to (\ref{kernel}). If we study the electric sector of
the theory, we find, because of (\ref{conn1}),
\begin{equation}\label{D1}
{\cal H}^{\hat{0}B}(\tau)=\eta^{\hat{0}\hat{0}}\eta^{BD}\left[
F_{\hat{0}D}(\tau)- c \int_{\tau_0}^{\tau}\left(
\Gamma_{0\hat{0}}{}^{C}F_{CD}+\Gamma_{0D}{}^{C}F_{\hat{0}C}
\right)\,d\tau'\right]\,
\end{equation} 
or
\begin{equation}\label{D2}
\boldsymbol{ D}=
\boldsymbol{E}+\int_{\tau_0}^{\tau}\left[{\boldsymbol{\omega}
(\tau-\tau')}\times\boldsymbol{E}(\tau')
-\frac{\boldsymbol{a}(\tau-\tau')}{c}\times\boldsymbol{B}(\tau')
\right]d\tau'\,.
\end{equation} 
Similarly, for the magnetic sector, the corresponding relations read
\begin{multline}\label{H1}
{\cal H}^{AB}= \eta^{AD}\eta^{BE}\Big[
F_{DE} \\ - c \int_{\tau_0}^{\tau}\left(
\Gamma_{0D}{}^{\hat{0}}F_{\hat{0}E}+\Gamma_{0D}{}^{C}F_{CE}+
\Gamma_{0E}{}^{\hat{0}}F_{D\hat{0}}+\Gamma_{0E}{}^{C}F_{DC}
\right)\,d\tau'\Big]\,
\end{multline}
or
\begin{equation}\label{H2}
\boldsymbol{ H}=
\boldsymbol{B}+\int_{\tau_0}^{\tau}\left[{\boldsymbol{\omega}
(\tau-\tau')}\times\boldsymbol{B}(\tau')+
\frac{\boldsymbol{a}(\tau-\tau')}{c}\times\boldsymbol{E}(\tau')
\right]d\tau'\,,
\end{equation}
respectively. Clearly, for {\em constant} $\boldsymbol{a}$ and
$\boldsymbol{\omega}$ our nonlocal relations (\ref{D2}) and
(\ref{H2}) are the same as \eqref{explicit1} and \eqref{explicit2}
provided we identify $\mathcal{H}$ with $\mathcal{F}$, i.e.\ we
postulate that the field actually measured by the accelerated observer
is the excitation $\mathcal{H}$. This agreement does not extend to the
case of \emph{non}uniform acceleration, however, as will be
demonstrated in the next section. 

\section{Nonuniform acceleration}
To show that the new ansatz \eqref{kernel} is different from
Mashhoon's ansatz \eqref{non-local1} for the case of nonuniform
acceleration even when we identify $\mathcal{H}$ with $\mathcal{F}$,
we proceed via contradiction. That is, let us assume that
$\mathcal{F}_{\alpha\beta} = \mathcal{H}_{\alpha\beta}$ and hence from
\eqref{D2} and \eqref{H2} 
\begin{equation}\label{nonuniform1}
K(\tau) = \begin{bmatrix} K_{\boldsymbol{\omega}} &
-K_{\boldsymbol{a}} \\  K_{\boldsymbol{a}}  &   K_{\boldsymbol{\omega}} 
\end{bmatrix} \;, 
\end{equation}
where $K_{\boldsymbol{\omega}} = \boldsymbol{\omega}(\tau) \cdot
\boldsymbol{I}$ and $K_{\boldsymbol{a}} = \boldsymbol{a}(\tau) \cdot
\boldsymbol{I}/c$. Here $I_A$, $(I_A)_{BC} = - \epsilon_{ABC}$, is a
$3\times 3$ matrix that is proportional to the operator of
infinitesimal rotations about the $e_{{A}}$-axis. We must now
prove that in general $R(\tau)$ given by \eqref{resolvent} cannot be
the resolvent kernel corresponding to $K(\tau)$ given by
\eqref{nonuniform1}. 

To this end, consider an observer that is accelerated at $\tau_0=0$
and note that for kernels of Faltung type in equations
\eqref{non-local1} and \eqref{non-local2} we can write
\begin{equation}
\overline{\mathcal{F}} = (I+\overline{K}) \overline{\hat{F}}
\qquad{\rm and}\qquad \overline{\hat{F}} =
(I+\overline{R})\overline{\mathcal{F}}\,,
\end{equation} respectively, where
$\overline{f}(s)$ is the Laplace transform of $f(\tau)$ defined by
\begin{equation}
\overline{f}(s) :=\int_0^\infty f(\tau)e^{-s\tau}
\,d\tau\end{equation} and $I$ is the unit $6\times 6$ matrix. Hence,
the relation between $K$ and $R$ may be expressed as
\begin{equation}\label{kernel-cond}
(I+\overline{K}) (I + \overline{R}) = I \;.
\end{equation}

\begin{center}
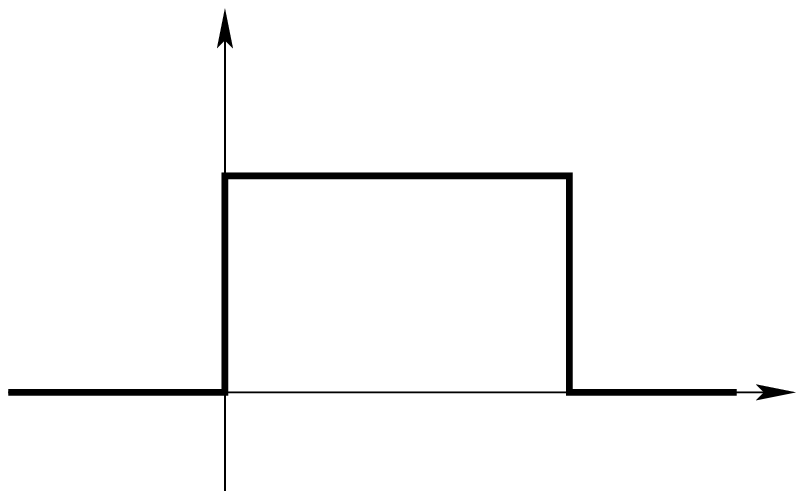 
\end{center}

\noindent Fig.2. The acceleration of an observer that is uniformly
accelerated only during a finite interval from $\tau=0$ to
$\tau=\alpha$. 
\bigskip

Imagine now an observer that is at rest on the $z$-axis for $-\infty <
\tau < 0$ and undergoes linear acceleration along the $z$-axis at
$\tau=0$ such that ${a}(\tau) = {g} > 0$ for $0\le
\tau<\alpha$ and ${a}(\tau)=0$ for $\tau\ge \alpha$ (see
Fig.~2). That is, the acceleration is turned off at $\tau=\alpha$ and
thereafter the observer moves with uniform speed $c\tanh({g}\alpha/c)$
along the $z$-axis to infinity. Thus in \eqref{nonuniform1},
$K_{\boldsymbol{\omega}}=0$ and $K_{\boldsymbol{a}} =
{a}(\tau) \, I_3 /c$. On the other hand, one can show that
\eqref{resolvent} can be expressed in this case as 
\begin{equation}
R(\tau) = {a}(\tau) \begin{bmatrix} U & V \\ -V & U 
\end{bmatrix} \;,
\end{equation}
where $U = J_3 \,\sinh\Theta$, $ V=I_3 \,\cosh\Theta$, and $(J_3)_{AB} =
\delta_{AB} - \delta_{A3}\delta_{B3}$. Here we have set $c=1$ and 
\begin{equation}
\Theta(\tau) = \int\limits_0^\tau {a}(\tau)\,d\tau = \begin{cases}
g \,\tau, & 0 \le \tau < \alpha, \\
g \,\alpha, & \tau \ge \alpha .
\end{cases}
\end{equation}
It is now possible to work out \eqref{kernel-cond} explicitly and 
conclude that for 
\begin{equation}
X(s) := \overline{{a}(\tau) \sinh\Theta}\;, \quad 
Y(s) := \overline{{a}(\tau) \cosh\Theta}\;, \quad 
Z(s) := \overline{{a}(\tau)} \;,
\end{equation}
we must have 
\begin{equation}
X = YZ\;, \quad Y=Z(1+X)\;. 
\end{equation}
These relations imply that 
\begin{equation}\label{ys0}
Y(s) = \frac{Z(s)}{1-Z^2(s)}\;. 
\end{equation}
On the other hand, we have
\begin{equation}
Z(s) = \int\limits_0^\infty {a} (\tau) e^{-s\tau} \, d\tau
= \frac{ {g} }{s} \left( 1 - e ^{-\alpha s} \right) \; 
\end{equation}
and 
\begin{eqnarray}\label{ys}
Y(s)& = &\frac{1}{2} \int\limits_0^\infty
{a}(\tau)\left(e^\Theta + e^{-\Theta} \right) e^{-s\tau}
\,d\tau\nonumber\\& =& \frac{{g}}{2} \left[ \frac{1 -
e^{-(s-{g})\alpha}}{s-{g}} + \frac{1 -
e^{-(s+{g})\alpha}}{s+{g}} \right]\,.
\end{eqnarray}
We consider only the region $s>g$ in which $X(s)$ and $Y(s)$ remain
finite for $\alpha\rightarrow\infty$. Comparing (\ref{ys}) with
\begin{equation}
 \frac{Z}{1-Z^2} = \frac{{g} s (1-e^{-\alpha s} )}{s^2
- {g}^2 (1-e^{-\alpha s})^2}\;,
\end{equation}
we find that, contrary to (\ref{ys0}), they do not agree except in the
$\alpha \rightarrow \infty$ limit (see Fig.3).  Therefore, we conclude
that the two models are different if one considers
\emph{arbitrary} accelerations.

\includegraphics[width=8cm,angle=-90]{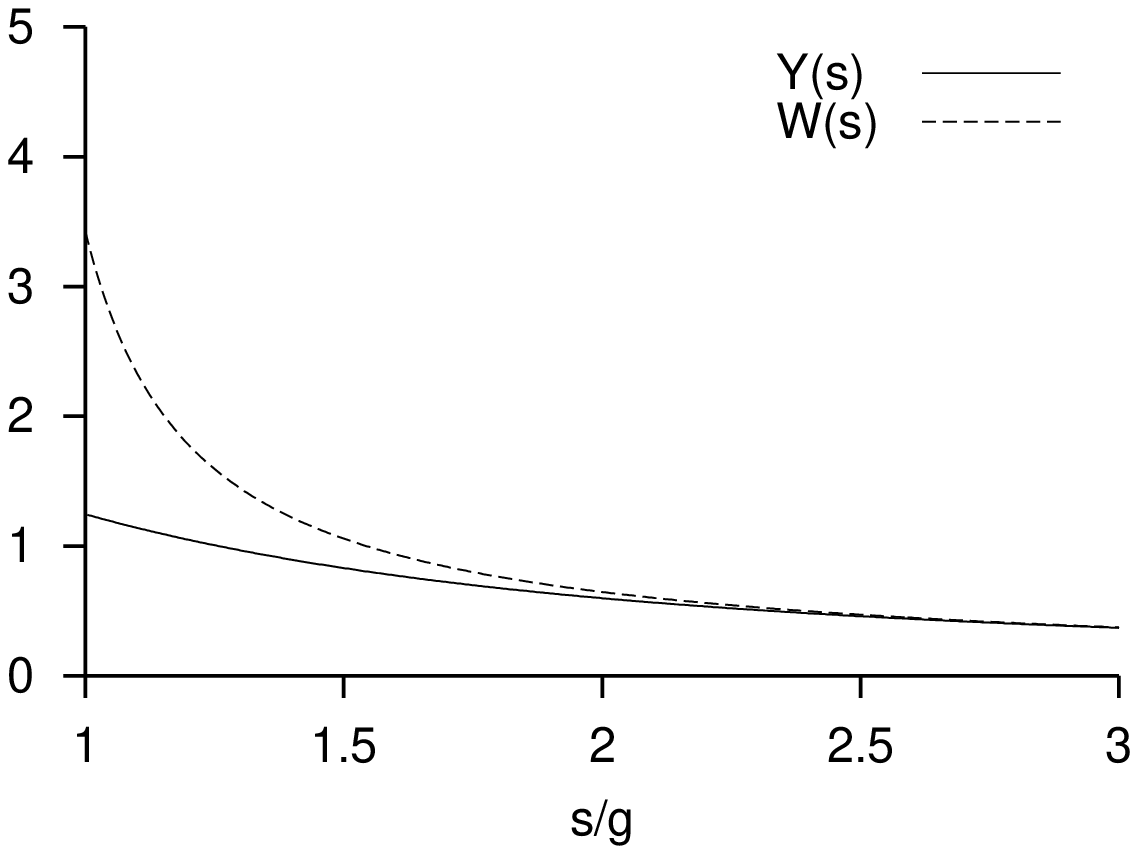} 
\bigskip

\noindent Fig.3. Plot of the functions $Y(s)$ and
$W(s):=Z(s)/[1-Z^2(s)]$ for $\alpha g=2$.
\bigskip 

\section{Discussion}
If one rewrites Mashhoon's nonlocal electrodynamics in the framework
of charge \& flux electrodynamics in vacuum by substituting the
generalization of equation \eqref{non-local2} for a congruence of
accelerated observers in equations \eqref{Maxwell}--\eqref{vacuum},
one finds a rather complicated implicit nonlocal constitutive law. The
Maxwell equations expressed in terms of the excitations $(\boldsymbol{
D},\boldsymbol{ H})$ and field strengths
$(\boldsymbol{E},\boldsymbol{B})$ remain the same, a fact which is
significant since otherwise the conservation laws of electric charge
and magnetic flux would be violated.

In this paper, we have developed an alternative nonlocal constitutive
ansatz within the framework of charge \& flux electrodynamics such
that the nonlocality is induced by the acceleration of the observer in
a similar way as in Mashhoon's model.

An explicit example of nonuniform acceleration has been used to show
that the two nonlocal prescriptions discussed here are in general
different. 

\bigskip

\noindent{\bf Acknowledgments:} One of the authors (FWH) would like to
thank Bahram Mashhoon and the Department of Physics \& Astronomy of
the University of Missouri-Columbia for the invitation for a two-month
stay and for the hospitality extended to him. This stay has been
supported by the VW-Foundation of Hannover, Germany. We are grateful
to Yuri Obukhov (Moscow) for highly interesting and critical
discussions on nonlocal electrodynamics. Thanks are also due to
Guillermo Rubilar (Cologne) for helpful correspondence.

\end{document}